\documentclass{elsart}

\usepackage{graphicx}

\usepackage{amsmath}
\usepackage{amssymb}

\begin{document}

\begin{frontmatter}



\title{Formation of hydrogen peroxide and water from the reaction of
 cold hydrogen atoms with solid oxygen at 10 K}



\author[label1]{N. Miyauchi\corauthref{cor1}}\ead{naoya@lowtem.hokudai.ac.jp},
 \author[label1]{H. Hidaka}, \author[label1]{T. Chigai},
 \author[label1]{A. Nagaoka}, \author[label1]{N. Watanabe},
 \author[label1]{A. Kouchi}

\corauth[cor1]{Corresponding author address.  Institute of Low
Temperature Science, Hokkaido University, N19-W8, Kita-ku, Sapporo
060-0819, Japan. FAX: +81-11-706-7142}
\address[label1]{Institute of Low Temperature Science, Hokkaido University, 
N19-W8, Kita-ku, Sapporo 060-0819, Japan}

\begin{abstract}
The reactions of cold H atoms with solid O$_2$ molecules were
 investigated at 10~K. The formation of H$_2$O$_2$ and H$_2$O has been
 confirmed by {\textit {in-situ}} infrared spectroscopy. We found that the
 reaction proceeds very efficiently and obtained the effective reaction
 rates. This is the first clear experimental evidence of the formation
 of water molecules under conditions mimicking those found in cold
 interstellar molecular clouds. Based on the experimental results, we
 discuss the reaction mechanism and astrophysical implications.

\end{abstract}

\end{frontmatter}


\section{Introduction}
Water is the most abundant solid material in space, and has been
observed in various astrophysical environments, such as outer planets,
satellites, comets, and interstellar clouds~\cite{schmitt1998}. Since
the solar system evolved from an interstellar molecular cloud, icy
objects in the solar system originated from the water ice formed in the
interstellar molecular cloud. Therefore, gaining an understanding of the
origin of water molecules in interstellar molecular clouds is critical
not only for discussing the origin of the solar system, but also for
understanding chemical evolution and the origin of
life~\cite{chela1996}. However, the formation mechanism of water
molecules in the interstellar clouds has not been understood to date. It
has been clarified that the formation of water molecules in the gas
phase is incapable of explaining the observed abundance in molecular
clouds~\cite{dHendecourt1985,hasegawa1992}. Thus, it has been suggested
that water molecules are synthesized by atomic reactions involving H and
O on pre-existing silicate or carbonaceous grains at around
10~K~\cite{dHendecourt1985,hasegawa1992,Tielens2005}.
\begin{figure}[htbp]
 \begin{center}
  \includegraphics[width=70mm]{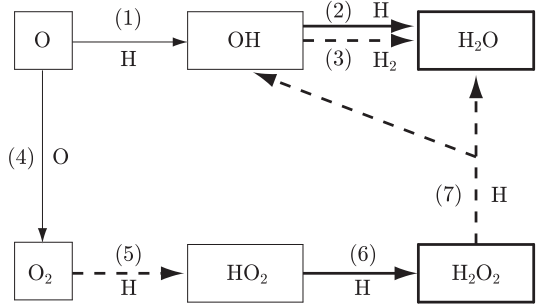} \caption{Possible reaction
scheme for formation of H$_2$O involving H and O atoms in interstellar
clouds (modified from~\cite{Tielens2005,tielens1982}). Atoms or
molecules and numerals next to the arrows indicate the reactants in the
reaction and the number of the reaction in the text,
respectively. Broken and solid arrows indicate reactions with and
without activation barriers, respectively. Thick and thin arrows denote
reactions investigated and not investigated in the present study,
respectively. Molecules observed in the present study are enclosed in
thick squares.} 
\label{fig:1}
 \end{center}
\end{figure}

The first possible process for H$_2$O formation is the sequential
hydrogenation of O atoms on grains:
\begin{equation}
 \mathrm{O} +  \mathrm{H}  \rightarrow \mathrm{OH},\label{eq:O+H} 
\end{equation}
\begin{equation}
 \mathrm{OH} +  \mathrm{H} \rightarrow  \mathrm{H}_2\mathrm{O}.\label{eq:OH+H} 
\end{equation}
These reactions have no activation barriers~\cite{herbst2007} The second possible process is the reaction
 of OH with hydrogen molecules absorbed on the surface of grains:
\begin{equation}
 \mathrm{OH} + \mathrm{H}_2 \rightarrow \mathrm{H}_2\mathrm{O} + \mathrm{H},\label{eq:OH+H2}
\end{equation}
which has an activation energy of 5.16~kcal/mol in the gas
phase\cite{schiff1973}. For the third process, Tielens and
Hagen\cite{tielens1982} proposed alternative reactions:
\begin{equation}
 \mathrm{O} + \mathrm{O} \rightarrow \mathrm{O}_2,\label{eq:O+O} 
\end{equation}
\begin{equation}
 \mathrm{O}_2 + \mathrm{H} \rightarrow \mathrm{HO}_2,\label{eq:O2+H} 
\end{equation}
\begin{equation}
 \mathrm{HO}_2 + \mathrm{H} \rightarrow \mathrm{H}_2\mathrm{O}_2,\label{eq:HO2+H}
\end{equation}
\begin{equation}
  \mathrm{H}_2\mathrm{O}_2 + \mathrm{H} \rightarrow \mathrm{H}_2\mathrm{O} + \mathrm{OH}.\label{eq:H2O2+H}
\end{equation}

Reaction (\ref{eq:O2+H}) has essentially no barrier; theoretically
estimated activation energies lie between 0.1 and
0.4~kcal/mol~\cite{Walch1988}. Reaction (\ref{eq:H2O2+H}) has activation
energies of 3.6-4.3~kcal/mol~\cite{Koussa2006}. All activation energies
referred here are value in the gas phase. There has been no data
available on a surface. Fig.~\ref{fig:1} summarizes the possible water
formation routes from reactions
(\ref{eq:O+H})-(\ref{eq:H2O2+H}). Cuppen and Herbst~\cite{herbst2007}
simulated the formation of water molecules in various environments of
interstellar clouds, and found that reactions (\ref{eq:O+H}) and
(\ref{eq:OH+H}) are the main routes in cool diffuse clouds and reactions
from (\ref{eq:OH+H2})- (\ref{eq:H2O2+H}) are the main routes in cold
molecular clouds. H and O atoms are the major gas-phase species in the
former clouds, whereas H$_2$ molecules are the major gas-phase species
in the latter clouds.

Hiraoka et al.~\cite{hiraoka1998} investigated reactions (\ref{eq:O+H})
and (\ref{eq:OH+H}) by spraying D atoms onto O atoms trapped in an
N$_2$O matrix at 12 K. They analyzed the products by using
temperature-programmed desorption (TPD) spectroscopy and observed
D$_2$O. A similar experiment has been performed by Dulieu et
al.~\cite{Dulieu2007} although their results are still very
preliminary. However, in the experiments of both groups, it remains
unclear whether D$_2$O is formed at 10 - 12 K or during heating. It is
essential to perform \textit{in-situ} analysis during H/D atom
irradiation to confirm the formation temperature of H$_2$O and to obtain
kinetic data. For reactions (\ref{eq:O2+H})-(\ref{eq:H2O2+H}), Klein
and Scheer~\cite{Klein1959} performed pioneering experiments on the
reaction of H atoms with solid O$_2$ at 20~K. Although they speculated
on the production of H$_2$O$_2$ and H$_2$O based on their experimental
results, they were unable to verify the production of these molecules
since no analysis of the products was performed. To investigate the
formation mechanism of water molecules in interstellar molecular clouds,
we focus on reactions (\ref{eq:O2+H})-(\ref{eq:H2O2+H}) and the
subsequent reactions (\ref{eq:OH+H}) and (\ref{eq:OH+H2}), and performed
H/D addition experiments to solid O$_2$ at 10~K.

\section{Experimental}
Experiments were performed using the Apparatus for SUrface Reaction in
Astrophysics (ASURA) system described
previously~\cite{watanabe2004,watanabe2006}. Briefly, solid O$_2$ with
an 8 monolayer (ML) thickness was produced by vapor deposition on an
aluminum substrate at 10~K at a deposition rate of 25 pm s$^{-1}$ in an
ultrahigh vacuum chamber at a pressure of about 10$^{-8}$ Pa. Atomic H
and D were produced by a microwave-induced plasma in a Pyrex tube and
transferred via a series of PTFE and aluminum tubes to the target.
The dissociation fraction was about 20\% at least. The atomic beams
could be cooled to 20~K in the aluminum tube that was connected to an He
refrigerator. The flux of atoms was controlled by varying the
temperature of the aluminum tube and measured by the method of Hidaka et
al.~\cite{hidaka2007}. The fluxes of both the H and D atoms were
2$\times$10$^{14}$~cm$^{-2}$~s$^{-1}$ in the present
experiments. Infrared absorption spectra of the sample solid during
irradiation by atoms were measured {\textit {in-situ}} by Fourier
transform infrared spectroscopy with a resolution of 4~cm$^{-1}$. After
irradiation, TPD spectra of the sample were obtained using a quadrupole
mass spectrometer at a heating rate of 4~K~min$^{-1}$.

\section{Results and discussion}
\begin{figure}[htbp]
 \begin{center}
  \includegraphics[width=90mm]{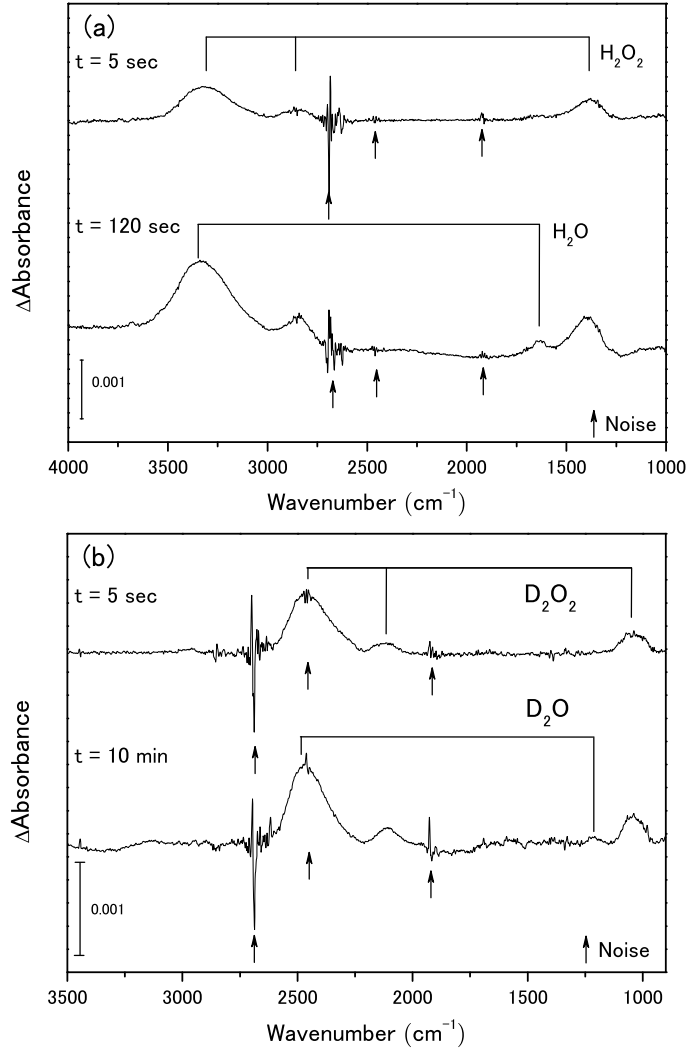}
 \caption{Spectral change in
 solid O$_2$ due to (a) cold H exposure and (b) D exposure. The spectra
 were obtained by subtracting the initial absorption spectra of
 non-irradiated samples from the spectra of H(D)-irradiated
 samples. Lines indicate newly formed molecules. Spikes with arrows
 indicate noise caused by vibration of the He refrigerator.}
 \end{center}
\end{figure}

Fig.~2a shows the change in a typical absorption spectra on irradiation
by 70-K-hydrogen atoms at 10~K. This figure shows the absorbance
variations from the initial spectrum of the sample, so that bands
appearing above the baseline indicate an increase in
absorbance. H$_2$O$_2$ appears immediately with bands at 3250, 2830, and
1405~cm$^{-1}$, then H$_2$O follows at 3432, 1650, and 820~cm$^{-1}$. No
product was observed when solid O$_2$ was exposed to an H$_2$ beam at 10
K, or when H atoms were sprayed onto the Al substrate without solid
O$_2$, demonstrating that the formation of H$_2$O$_2$ and H$_2$O in the
present experiments is due to H atom irradiation onto O$_2$. No
intermediate radicals (e.g., HO$_2$) were observed, suggesting that the
reaction rate of (\ref{eq:O2+H}) is much slower than that of
(\ref{eq:HO2+H}). In this experiment, the H flux is sufficiently large
so that the reaction (\ref{eq:O2+H}) immediately proceeds to
produce HO$_2$. The formation of H$_2$O$_2$ and H$_2$O were
also observed when solid O$_2$ was exposed to an H beam at temperatures
between 15 K and 20 K. Fig.~2b clearly reveals the production of
D$_2$O$_2$ and D$_2$O in D-exposure experiments, confirming that the
H$_2$O produced in the present experiments is not a contaminant. The
production of H$_2$O$_2$(D$_2$O$_2$) and H$_2$O(D$_2$O) was also
indicated in TPD spectra obtained after H(D)-exposure.
\begin{figure}[htbp]
 \begin{center}
  \includegraphics[width=85mm]{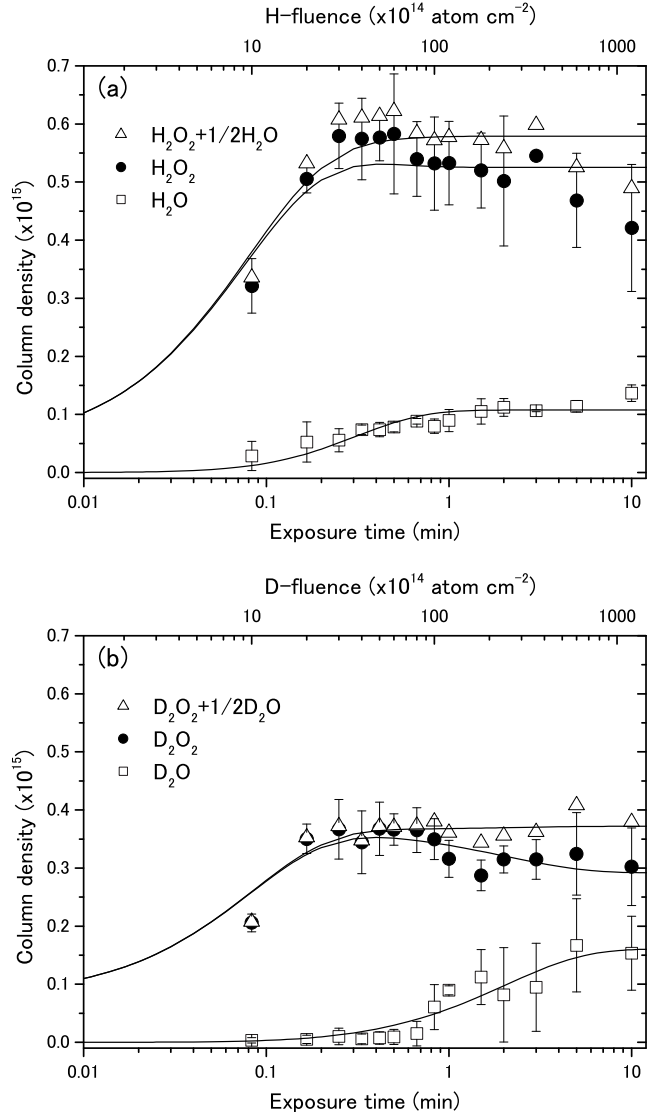}
 \caption{ Variation in column densities of products (a) for H$_2$O$_2$
  and H$_2$O with H exposure and (b) for D$_2$O$_2$ and D$_2$O with D
  exposure. Each data point represents the average of three
  measurements, and error bars represent the statistical errors. The
  upper abscissa indicates the fluence of H(D) atoms. The solid lines
  are fitted to the curves defined by Eqs. (\ref{eq:chigai1c}) to
  (\ref{eq:chigai2b}). The amount of O$_2$ reduction is assumed to be
  equal to the sum of the increase in H$_2$O$_2$ (D$_2$O$_2$) and
  1/2H$_2$O (1/2D$_2$O). Fitting curves of H$_2$O and D$_2$O shown are
  two times larger than those obtained from Eq. (\ref{eq:chigai2b}).}
 \end{center}
\end{figure}

The amount of products are calculated and plotted as a function of
exposure time in Fig.~3. These were obtained from the band areas in the
spectra and the reported and estimated integrated band strengths. The
integrated band strengths used are 1.2$\times$10$^{-17}$ and
2.1$\times$10$^{-17}$ molecule~cm$^{-1}$ for H$_2$O
(1655~cm$^{-1}$)~\cite{gerakins1995} and H$_2$O$_2$
(1347~cm$^{-1}$)~\cite{loeffler2006}, respectively. Since no integrated
band strengths for D$_2$O at 1200~cm$^{-1}$ and D$_2$O$_2$ at
1033~cm$^{-1}$ have been reported, we estimated these in the following
manner. Bergren et al.~\cite{bergren} measured the optical constants of
OH(OD)-stretching modes in amorphous H$_2$O and D$_2$O at 3200~cm$^{-1}$
and 2400~cm$^{-1}$, respectively. Assuming that the ratio of integrated
band strengths for H$_2$O and D$_2$O in OH(OD)-stretching modes are the
same as those in OH(OD)-bending modes, we estimated the integrated band
strength of OD-bending modes for D$_2$O (1200~cm$^{-1}$) to be
8$\times$10$^{-18}$ molecule~cm$^{-1}$. No data is available for
D$_2$O$_2$, not even optical constants. So we assumed that the ratio of
integrated band strengths for H$_2$O/D$_2$O$ = 0.67$ is the same as that
for H$_2$O$_2$/D$_2$O$_2$, and estimated the integrated band strength of
OD-bending modes for D$_2$O$_2$ (1033~cm$^{-1}$) to be
1.5$\times$10$^{-17}$ molecule~cm$^{-1}$. This result is consistent with
the spectral features of OH and OD-bending modes of amorphous H$_2$O$_2$
and D$_2$O$_2$ measured by Lannon et al.~\cite{lannon1971}, which show
that the ratio of peak heights for H$_2$O$_2$/D$_2$O$_2$ is 0.75.

Fig.~3a clearly shows that the formation of H$_2$O$_2$ and H$_2$O is
very rapid and efficient; H$_2$O$_2$ and H$_2$O are observed even after
exposure for 5 s. When H atoms are irradiated onto solid CO at 10
K~\cite{watanabe2004}, the production of H$_2$CO and CH$_3$OH are
observed after 30~s and 1~min, respectively. The present experimental
results confirm that reactions (\ref{eq:O2+H})-(\ref{eq:H2O2+H}), which
were initially proposed based on theoretical considerations
\cite{tielens1982,herbst2007}, proceed at 10~K. Since reaction
(\ref{eq:O2+H}) essentially has no barrier~\cite{Walch1988}, it is
natural that this reaction proceeds very rapidly. While reaction
(\ref{eq:H2O2+H}) has an activation energy of
3.6-4.3~kcal/mol~\cite{Koussa2006}, it has been suggested that this
reaction does not proceed by an Arrhenius-type reaction at 10~K, but
rather proceeds by a tunneling reaction even at 10~K. As described
later, isotope effect (H or D) on reaction (\ref{eq:H2O2+H}) has been
observed, strongly supporting that reaction (\ref{eq:H2O2+H}) proceeds
by the tunneling reaction. After the formation of OH by reaction
(\ref{eq:H2O2+H}), there are two possible reaction routes for further
H$_2$O formation: reactions (\ref{eq:OH+H}) and (\ref{eq:OH+H2}). This
is because the H-beams used in the present experiments do not consist of
pure H atoms, but rather are a mixture of H and H$_2$. Reaction
(\ref{eq:OH+H}) has no activation barrier, while the activation energy
of reaction (\ref{eq:OH+H2}) is 5.16~kcal/mol~\cite{schiff1973}
for reaction (\ref{eq:OH+H2}). It seems probable that H$_2$O is
formed by reaction (\ref{eq:OH+H}), although we currently have no direct
evidence for this. Fig.~3b shows that D$_2$O$_2$ and D$_2$O are produced
very rapidly and efficiently in the D-exposure experiments, although the
formation of D$_2$O is slightly slower than in H-exposure experiments.

In the present experiments, it is impossible to directly measure the
decrease in O$_2$, since O$_2$ is infrared inactive. Thus, it is
reasonable to assume that the summation of 0.5H$_2$O and H$_2$O$_2$ is
equal to the reduction of O$_2$. The O$_2$ loss becomes saturated after
long exposure. This type of saturation was also found previously for the
successive hydrogenation of CO at 8-15 K by Watanabe et
al.~\cite{watanabe2004,watanabe2003}. They concluded from the dependence
of the saturation value on the sample thickness that the saturation is
related not to the balance between the forward and reverse reactions,
but rather to the diffusion length of H in solid CO and CO-H$_2$O mixed
ice. This suggests that in the present experiments the reaction occurs
only at the surface. The structure of the molecular layer may thus be
onion-like, with H$_2$O on the top, H$_2$O$_2$ in the middle, and O$_2$
at the bottom. The amount of H$_2$O$_2$ is greater than that of
D$_2$O$_2$, suggesting that the diffusion length of H into O$_2$ is
longer than that of D into O$_2$. On the other hand, the amount of
H$_2$O is the same as that of D$_2$O, suggesting that the water
formation reaction occurs only at surface. The small amount of observed
H$_2$O compared to H$_2$O$_2$ may be attributed to desorption of
products caused by the release of the heat of reaction.

To discuss the reaction mechanism based on the isotope effect, we next
derive effective reaction rate constants by fitting the data in
Fig.~3. To simplify the problem, we assume that reactions
(\ref{eq:O2+H})-(\ref{eq:H2O2+H}), (\ref{eq:OH+H}) and (\ref{eq:OH+H2})
can be simplified to the following consecutive reaction:
\begin{equation}
 \mathrm{O}_2 \xrightarrow{k_1} \mathrm{H}_2\mathrm{O}_2 \xrightarrow{k_2} 2\mathrm{H}_2\mathrm{O}\label{eq:k1k2}
\end{equation}
where $k_i$ is the reaction rate constant of the $i$th reaction. Since,
it is very difficult to measure the surface density of H atoms,
$n_\textrm{H}$, in the present study, we were not able to obtain $k_i$
directly. Instead, we define effective rate constants :
$k'_{\textrm{H}i} = k_{i}n_\textrm{H}$. Since $n_\textrm{H}$ is
considerd to be time-independent, these reactions are assumed to be
pseudo-first-order reactions.

Furthermore, we assume that the products have a layered structure, as
mentioned above. In the deepest layer, no reaction of O$_2$ with H atoms
occurs. In the middle layer that has a thickness of $a$, the reaction
from O$_2$ to H$_2$O$_2$ occurs, but that from H$_2$O$_2$ to H$_2$O does
not occur. In the uppermost layer with a thickness of $b$, the entire
reaction (\ref{eq:k1k2}) occurs. The rate equations for the number
densities $n_i$ of O$_2$, H$_2$O$_2$ and H$_2$O are given by
\begin{equation}
 \frac{dn_0}{dt} = -k'_{\textrm{H}1} n_0,\label{eq:chigai1a}
\end{equation}
\begin{equation}
 \frac{dn_1}{dt} = k'_{\textrm{H}1} n_0 -k'_{\textrm{H}2} n_1,\label{eq:chigai1b}
\end{equation}
\begin{equation}
 \frac{dn_2}{dt} = 2k'_{\textrm{H}2} n_1.\label{eq:chigai1c}
\end{equation}
The solutions of equations (\ref{eq:chigai1a})-~(\ref{eq:chigai1c}) for
the initial conditions of $n_0(t=0) = n_0^0$, and $n_1 (0) = n_2 (0) =
0$ are given by
\begin{equation}
 \frac{n_0}{n_0^0} = b\mathrm{e}^{-k'_{\textrm{H}1} t} + a\mathrm{e}^{-k'_{\textrm{H}1} t},\label{eq:chigai2a}
\end{equation}
\begin{equation}
 \frac{n_1}{n_0^0} = b\frac{k'_{\textrm{H}1}}{k'_{\textrm{H}1} - k'_{\textrm{H}2}}\left(\mathrm{e}^{-k'_{\textrm{H}2} t} -
					    \mathrm{e}^{-k'_{\textrm{H}1} t}\right) +
 a\left(1-\mathrm{e}^{-k'_{\textrm{H}1} t}\right),\label{eq:chigai2b}
\end{equation}
\begin{equation}
 \frac{n_2 /2}{n_0^0} = b\left[1+\frac{1}{k'_{\textrm{H}1} - k'_{\textrm{H}2}}\left(k'_{\textrm{H}2}
						   \mathrm{e}^{-k'_{\textrm{H}1} t} -
						  k'_{\textrm{H}1} \mathrm{e}^{-k'_{\textrm{H}2} t}\right)\right],\label{eq:chigai2c}
\end{equation}
where $n_0 +n_1 + n_2/2 = n_0^0$.  In the case of D exposure, the same
reaction as that for H exposure is assumed to occur, namely,
\begin{equation}
 \mathrm{O}_2 \xrightarrow{k'_{\textrm{D}1}} \mathrm{D}_2\mathrm{O}_2 \xrightarrow{k'_{\textrm{D}2}} \mathrm{D}_2\mathrm{O},\label{eq:k3k4}
\end{equation}

where $k'_{\textrm{D}i} = k_in_\textrm{D}$, and $n_\textrm{D}$ is the
surface density of D atoms.  Effective rate constants, $k'_{\textrm{H}i}$ and
$k'_{\textrm{D}i}$, obtained by fitting the data are shown in Table 1 together
with those for the first steps of successive hydrogenation (deuteration)
of CO~\cite{watanabe2006,hidaka2007}:
\begin{equation}
\mathrm{CO} \xrightarrow{k'_{\textrm{H}3}} \mathrm{HCO} \rightarrow
 \mathrm{H}_2\mathrm{CO} \rightarrow \mathrm{CH}_3\mathrm{O} \rightarrow \mathrm{CH}_3\mathrm{OH},\label{eq:CO+H}
\end{equation}
\begin{equation}
\mathrm{CO} \xrightarrow{k'_{\textrm{D}3}} \mathrm{DCO} \rightarrow
 \mathrm{D}_2\mathrm{CO} \rightarrow \mathrm{DH}_3\mathrm{O} \rightarrow \mathrm{DH}_3\mathrm{OD}.\label{eq:CO+D}
\end{equation}
Using the relation of $n_\textrm{H}/n_\textrm{D}=1$~\cite{hidaka2007},
the ratios of the reaction rate constants are estimated:
$k'_{\textrm{H}1} /k'_{\textrm{D}1} = 1$ and $k'_{\textrm{H}2} /
k'_{\textrm{D}2} = 8$. The former result states that there is no
difference in rate constants between O$_2$ + H and O$_2$ + D at 10~K,
and it is consistent with the fact that reaction (\ref{eq:O2+H}) has
essentially no barrier~\cite{Walch1988}. The latter result for the ratio
of reaction rate constants between H$_2$O$_2$ + H and D$_2$O$_2$ + D
($k'_{\textrm{H}2} / k'_{\textrm{D}2} = 8$) is reasonable for a
tunneling reaction with an activation energy of 3.6-4.3~kcal/mol~\cite{Koussa2006}. In the case of CO + H(D), the ratio of reaction
rate constants, $k'_{\textrm{H}3} / k'_{\textrm{D}3}$, lies between 10
(from Table 1) and 13~\cite{hidaka2007} with an activation energy of
about 4~kcal/mol. In any case, these data are very useful for estimating the
barrier height and width of a potential, and potential energy surface as
discussed by Hidaka et al.~\cite{hidaka2007}. As already mentioned
qualitatively, the rate constants of (\ref{eq:k1k2}) and (\ref{eq:k3k4})
are one to two orders of magnitude greater than those of (\ref{eq:CO+H})
and (\ref{eq:CO+D}).
\begin{table}[htbp]
\begin{center}
\caption{Effective reaction rate constants (min$^{-1}$)}\label{tab:1}
\begin{tabular}{ccccccc} \hline
Temperature &  &  &Reaction  &  &  &  \\
(K) & O$_2$+H & H$_2$O$_2$+H & O$_2$+D & D$_2$O$_2$+D & CO+H & CO+D \\\cline{2-7}
 & $k'_{\textrm{H}1}$ & $k'_{\textrm{H}2}$ & $k'_{\textrm{D}1}$ & $k'_{\textrm{D}2}$ & $k'_{\textrm{H}3}$ & $k'_{\textrm{D}3}$ \\\hline
10 & 12.8 & 3.9 & 12.0 & 0.49 & 0.14$^a$ & 0.014$^b$ \\
15 &  &  &  &  & 0.41$^c$ & 0.033$^c$ \\\hline
\end{tabular}
\end{center}
a,b,c: obtained  with the H(D) flux of  1$\times$10$^{14}$cm$^{-2}$s$^{-1}$~\cite{watanabe2007}

a: for pure CO \cite{watanabe2006}; b: for pure CO \cite{hidaka2007}; c:
 for 0.8 ML CO on amorphous H$_2$O \cite{hidaka2007}

\end{table}

\section{Astrophysical implications}
Our results have several implications for ice in space. Assuming a
hydrogen number density of 1~cm$^{-3}$, the H fluences in a 10-K
molecular cloud will be 1.3$\times$10$^{16}$ and 1.3$\times$10$^{17}$
cm{$^{-2}$} over 10$^4$ and 10$^5$ years, respectively. In the present
experiment, these fluences approximately correspond to exposure times of
1 and 10 minutes. Considering that the lifetime of molecular clouds is
of the order of 10$^6$ years, it is reasonable to say that water
formation by reactions (\ref{eq:O2+H})-(\ref{eq:H2O2+H}) occurs very
quickly in molecular clouds. Based on computer simulations of surface
reaction networks, Tielens and Hagen~\cite{tielens1982} and Cuppen and
Herbst~\cite{herbst2007} conjectured that reactions
(\ref{eq:O+O})-(\ref{eq:H2O2+H}) occur in cold molecular clouds in which
the ratio H/H$_2$ is low. The present results strongly support their
conclusion experimentally, although the contribution of reactions
(\ref{eq:O+H})-(\ref{eq:OH+H2}) is still unclear.

Solid O$_2$ is expected to occur on the grain surface with H$_2$O and CO
ice in interstellar molecular clouds and to be observable by infrared
telescopes since the 6.44-$\mu$m O=O stretching mode is observable when
O$_2$ is surrounded by other molecules~\cite{ehrenfreund1992} and because
the 4.67-$\mu$m feature of CO is altered when O$_2$ is mixed with CO
\cite{vandenbussche1999}. However, no positive detection has been
reported so far~\cite{vandenbussche1999}. Present experiments clearly
explain why no O$_2$ has been observed: if an O$_2$ molecule is formed
on grain surface it will react very quickly with H to form H$_2$O$_2$
and H$_2$O within 10$^{4-5}$ years as shown in Figs. 3a.

Recent astronomical observations have revealed that the abundances of
some deuterated molecules are up to four orders of magnitude greater
than the cosmic D/H ratio of 1.5$\times$10$^{-5}$. In the case of water,
a HDO/H$_2$O ratio of 0.03 has been measured in a solar-type
protostar~\cite{paris2005b}. Many theoretical models, including pure gas
phase models~\cite{roberts2000} and gas-grain models~\cite{tielens1983},
have been proposed. However, only a few experimental studies have been
performed to verify the latter models~\cite{hidaka2007,nagaoka2005}. In
the case of formaldehyde and methanol, the successive addition of D to
CO (reaction~(\ref{eq:CO+D})) is not favorable for producing
deuterated formaldehyde and methanol~\cite{hidaka2007}. Instead, H-D
substitution in formaldehyde and methanol are necessary to achieve the
observed abundances of deuterated spiecies~\cite{nagaoka2005}. However,
Nagaoka et al.~\cite{nagaoka2005} found experimentally that no
deuteration of H$_2$O occurs by H-D substitution under D exposure at
10-20~K, even for fluences of up to
5$\times$10$^{18}$~cm$^{-2}$. Considering the ratios of
$k'_{\textrm{H}1} / k'_{\textrm{D}1} = 1$ and $k'_{\textrm{H}2}
/k'_{\textrm{D}2} = 8$ obtained in the present experiments and the D/H
atom ratio of 0.1 or less expected in molecular
clouds~\cite{roberts983}, deuterium addition to O$_2$ is favorable for
producing the observed amount of HDO; the observed HDO/H$_2$O ratio of
0.03 can be achieved in a time scale between 10$^4$ and 10$^5$
years. Although the discussion is indefinite at present, the
experimental results presented in this study provide a basis for
discussing H/D fractionation.

\vspace*{1ex}
\begin{flushleft}
 \textbf{Acknowledgements}
\end{flushleft}
\vspace*{1ex}
This work was partly supported by a Grant-in-Aid for Scientific Research
from the Japan Society for the Promotion of Science and the Ministry of
Education, Science, Sports, and Culture of Japan.

\end{document}